%% file: nu5gamma.tex
\input  macros.tex

\input epsf.tex

\nopageonenumber
\baselineskip = 18pt
\barsoff


\def\subsubsection#1{\medskip\noindent {\it #1} \medskip\noindent}

\def\lrderiv#1{{\leftrightarrow 
\atop {  }}  {\kern-0.8em \partial_#1} \,}
\def\lrdel{{\leftrightarrow 
\atop {  }}  {\kern-0.8em \nabla} \,}
\def\lrdelsl{{\leftrightarrow 
\atop {  }}  {\kern-0.8em \delsl} \,}
\def\pbar{\ol{p}}

\def\nubar{\ol{\nu}}
\def\ebar{\ol{e}}

\def\sw{s_w}

\def\mw{M_\ssw}
\def\mz{M_\ssz}
\def\eff{{\rm eff}}
\def\GF{G_\ssf}

\def\caption#1{\noindent {\baselineskip 10pt \eightrm #1}}


\line{hep-ph/0006165 \hfil McGill-00/18}

\title
\centerline{Two-Neutrino Five-Photon Scattering}
\centerline{at Low Energies}
\endtitle

\vskip 0.35in
\authors
\centerline{Y. Aghababaie and C.P. Burgess}
\vskip .07in
\centerline{\it ${}^a$ Physics Department, McGill
University} 
\centerline{\it 3600 University St., Montr\'eal,
Qu\'ebec,  Canada, H3A 2T8.} 
\endauthors

\abstract
We extend earlier constructions of the effective action for neutrino-photon 
scattering, using the connection between low-energy neutrino-photon and 
photon-photon scattering together with the known effective Lagrangian 
describing low-energy photon scattering in QED.  
We use this effective action to calculate analytic expressions for
the low-energy cross section for the (unpolarised) processes
$\nu \nubar \to 5\, \gamma$, $\nu \gamma \to \nu + 4\,\gamma$ and
$\gamma \gamma \to \nu \nubar + 3\,\gamma$. As a byproduct we derive
compact expressions for the $N$-body phase-space integrals for massless
particles, including those having non-trivial tensor-structure.
\endabstract
  

\section{Introduction}

Neutrinos and photons may be, with gravitons, the only particles which
are massless, or very nearly so. As such, they are the only degrees of
freedom which arise at extremely low energies within the vacuum sector
(and possibly within other sectors) of the Standard Model (SM). This
makes the study of their low-energy interactions a theoretical 
laboratory for very-low-energy Standard-Model physics. 
This study may also have practical applications, despite the extremely
weak strength of the interactions, because neutrinos play a unique role within 
the extreme environments found in astrophysics and cosmology. Precisely
because of their weak couplings they are often the mediators of 
dynamically interesting processes, for instance by being responsible for
heat and momentum transfer, especially in the late stages of stellar
collapse. 

\ref\Gell{M. Gell-Mann, \prl{6}{61}{70}.}
\ref\Dicussupp{D.A. Dicus, C. Kao and W. Repko, \
\prd{48}{93}{5106--5108} ({\tt hep-ph/9305284}).} 
\ref\Yang{C.N. Yang, \pr{77}{50}{242}.}
\ref\Dicuslow{D.A. Dicus, C. Kao and W. Repko, \
\prl{79}{97}{569--571} ({\tt hep-ph/9703210}).}
\ref\Dicushigh{D.A. Dicus, C. Kao and W. Repko, \
\prd{59}{99}{013012} ({\tt hep-ph/9806499});\
{\it ibid.} 013005 ({\tt hep-ph/9806499}). }
\ref\Matiasdirect{A. Abada, J. Matias and R. Pittau, \
\npb{543}{99}{255--268}, ({\tt hep-ph/9808294}).}
\ref\magnetic{See for example, D.A Dicus and W. Repko, \
{\tt hep-ph/0003305}, and references therein.}
\ref\giestwototwo{H. Gies and R. Shaisultanov, {\it Phys. Lett.} 
{\bf B480} (2000) 129--134.}

$2\to 2$ processes --- like $\nu \nubar \to \gamma\gamma$ and
$\nu \gamma \to \nu \gamma$ --- were first studied long ago and were 
found to be highly suppressed \Gell,\Dicussupp. The suppression 
arises because Yang's theorem \Yang\ prohibits the coupling of two photons to
a state of angular momentum one, and this ensures that the
$O(1/\mw^2)$ contribution to the amplitude for these $2\to 2$ 
process must be zero (but see \giestwototwo).
The dominant contribution therefore arises at $O(1/\mw^4)$, making
it smaller than the nominally negligible processes which arise at
higher order in $\alpha$, such as $\nu \nubar \to 3\gamma$ and
$\nu \gamma \to \nu + 2 \gamma$ \Dicuslow. This observation has
stimulated more detailed studies of these reactions (both at low
energies and at energies above the electron mass \Dicushigh,
\Matiasdirect), as well as generating searches for practical applications of 
these $2 \to 3$ processes, such as within stars or in neutrino-photon
scattering in the presence of magnetic fields \magnetic.

\ref\Matiasone{A. Abada, J. Matias and R. Pittau, \
\prd{59}{99}{013008} ({\tt hep-ph/9806383}).}

As pointed out in ref \Dicuslow, the effective action for these $2\to 3$ 
processes is related to the known Euler-Heisenberg effective
action for four-photon scattering through the replacement of one of 
the electromagnetic tensors, $F_{\alpha\beta}$, by a neutrino `field strength',
of the form $N_{\alpha\beta} = \partial_\alpha \, \Bigl( \nubar \gamma_\beta 
\; \gamma_\ssl 
\nu \Bigr) - (\alpha \leftrightarrow \beta)$. 
In ref. \Matiasone, this connection was examined in some detail:
a particular combination of Feynman diagrams was employed
to explicitly show the mapping between the calculation of
the Euler-Heisenberg action and that governing the interactions of
neutrino-antineutrino pairs with three photons.

In this note we have two goals. Our main new result is to extend this treatment to 
the next least complicated case: that of 2 neutrinos interacting with 5 photons.
We do so by computing the relevant terms of the low-energy effective neutrino-photon
lagrangian, and use them to calculate analytic expressions for the low-energy
neutrino-photon scattering cross sections. In the centre-of-mass (CM) 
we find these to be of order  $\sigma \sim {\alpha^3 \, \GF^2 \, E^{18} /
(2\pi)^6 m_e^{16}}$, where $E$ denotes the CM scattering energies, as
compared with the $2\to 2$ result: $\sigma \sim 200\, {\alpha \GF^4 E^6 / \pi^3}$.
In principle, the numerical factors are such that the $2\to 5$ processes can dominate the 
$2 \to 2$ processes for $E \sim m_e$ (which, of course, lies at the 
limit of validity of the low-energy approximation), although we know of no practical
application of this observation.

Our secondary goal is not so much new as it is explanatory. Preparatory to
describing the above results we recast the argument for the suppression of the
$2 \to 2$ processes into a more modern effective-lagrangian language. We also
rederive the connection between the electromagnetic and neutrino scattering
processes within this context. Although
these are old results, we hope that their recasting in this way may
suggest more applications elsewhere. 

Our presentation is as follows. In the next section, \S2, we review the low-energy
limit of neutrino-photon scattering, rederiving both the connection to the 
Euler-Heisenberg effective lagrangian and the suppression of $2 \to 2$ processes.
This is followed in \S3 by the derivation of the low-energy effective action for
$2 \to 5$ neutrino-photon interactions.  \S4 then applies this action to
compute the three $2 \to 5$ cross sections: $\nu \nubar \to 5 \gamma$,
$\nu \gamma \to \nu + 4 \gamma$ and $\gamma \gamma \to \nu \nubar + 3 \gamma$. 
We conclude in the last section, \S5, with comments and final remarks.  
Finally, an appendix describes an efficient method for computing the relevant $N$-body
phase-space integrals which are encountered.

\section{Low-Energy Neutrino-Photon Scattering Revisited}

Because the SM contains no direct (tree-level) couplings between neutrinos
and photons, the starting point for calculating their 
very-low-energy \foot\vle{We use the name `very-low-energy' to mean energies
below the electron mass, in order to distinguish this from other potential
notions of `low energy', such as $\ss E \ll \mw$.} interactions
is the SM description of their couplings to other particles. These other particles
then generate the effective neutrino-photon interactions once they are integrated
out to produce the very-low-energy theory. The effective couplings are nonrenormalizable,
in the sense that they are proportional to inverse powers of the masses of the
particles which were integrated out to obtain them. Our main interest in what
follows is in the dominant interactions at very low energy and so we focus
on integrating out the lightest particles. Since the two
lightest particles which couple to both neutrinos and photons are electrons
and muons, we concentrate our attention on these. 

\subsection{The Weak-Scale Effective Theory}
At energies below the $W$-boson mass, the couplings of neutrinos and photons
to charged leptons are described by the effective lagrangian obtained 
from the SM by integrating out the top quark and the electroweak gauge bosons, 
$W$ and $Z$. The resulting effective interactions which are of most 
interest in what follows are those which are suppressed by the fewest 
powers of $\mw$ or $\mz$. Those involving just neutrinos, photons and
charged leptons, obtained by matching 
to the SM at $\mu = \mw$, are given by:\foot\conventions{Like all God-fearing
people, our conventions are: $\ss \eta_{\mu\nu} = (-,+,+,+)$ and $\ol\psi = i \psi^\dagger \gamma^0$.} 
\eq
\label\FFGMlagr
\Scl_{\rm wk}(\mu = \mw) = e \, A_\mu \; J^\mu_{\rm em}
+ {\GF \over \sqrt 2} \; \sum_{klmn} 
\Bigl( i\nubar_k \gamma_\mu \Pl \nu_l \Bigr) \; L_{klmn}^\mu + O\left(
{1 \over \mw^4} \right), 
\eeq
where $k,l,m,n = e,\mu,\tau$ run over the three lepton flavours, 
and the charged-lepton currents are given by 
\eq
\label\currents
\eqalign{
L^\mu_{klmn} &= i\ol{\ell}_m \; \gamma^\mu \left( v_{klmn} +
a_{klmn} \gamma_5 \right) \ell_n ,\cr
J^\mu_{\rm em} &= - \sum_k i\ol{\ell}_k \; \gamma^\mu \; \ell_k .\cr}
\eeq
The effective couplings, $v_{klmn}$ and $a_{klmn}$, as found from tree-level
matching are given by:
\eq
\label\ecouplings
\eqalign{
v_{klmn}(\mu = \mw) &= \delta_{kn}\, \delta_{lm} + \delta_{kl}\, \delta_{mn}
\left( - \, \hf + 2\sw^2 \right)  \cr
\hbox{and} \qquad a_{klmn}(\mu = \mw) &= \delta_{kn}\, \delta_{lm} 
- \, \hf \, \delta_{kl} \, \delta_{mn}, \cr}
\eeq
at tree level, where $\sw = \sin\theta_w$ is the sine of the weak mixing 
angle. 

Radiative corrections are easily incorporated into this language. Those loops 
involving high energy degrees of freedom (those involving particles having 
masses as large as $\mw$ or larger) are included by matching to the SM with 
higher-loop accuracy. Intermediate-scale loops (involving particles having
masses between $m_e$ and $\mw$) are obtained by running the effective theory down
to each new particle threshold, and then matching across this threshold.

One such high-energy loop generates an effective coupling between neutrinos and photons which is proportional to $1/\mw^4$ \Dicuslow:
\eq
\label\nunuphph
\Scl_\eff(\mu) = {4 \alpha \over \pi \mw^2} \; \left( {\GF \over \sqrt2} \right)
\; \left[ 1 + {4 \over 3} \; \ln \left( { \mw^2 \over \mu^2} \right)
\right] \left( i\ol\nu \; \gamma_\alpha \gamma_\ssl \lrderiv{\beta} \nu \right)
\; F^{\beta \lambda} \, {F^\alpha}_\lambda .
\eeq
As we shall see, this particular higher-dimension interaction is {\it not}
generated when lighter particles are integrated out, and so it is the dominant
contribution to low-energy $2\to 2$ photon-neutrino scattering even though it
is suppressed by four powers of $\mw$.\foot\whatismu{In very-low-energy scattering
applications it is the effective couplings renormalized at the electron mass
which are required, so $\ss \mu = m_e$ is used in eq.~\nunuphph.}

Imagine now writing down the effective theory at scale $\mu = m_e$, just before 
integrating out the electron. The only particles in this low-energy theory are the
electron, photon and neutrinos. Conservation of electric charge and 
lepton numbers require the lowest-dimension neutrino couplings to electrons 
in this theory to again have the form of eq.~\FFGMlagr, although now
restricted to electrons and neutrinos. Furthermore, since all of the neutrino 
interactions in this effective theory must vanish in the 
limit where the $W$ and $Z$ become infinitely massive,
they must be proportional to at least one factor of $\GF$. 

It follows that the dominant low-energy interactions in the
effective theory at this scale can differ from the electron terms
of eq.~\FFGMlagr\ only through the values taken by the
the coefficients $v_{klee}(\mu = m_e)$ and $a_{klee}(\mu = m_e)$.
Happily enough, it also happens that $v_{klee}(\mu = m_e)$ cannot differ 
from its value, eq.~\ecouplings, at $\mu = \mw$, because the current
$\ebar \, \gamma^\alpha \, e$ is conserved, and so does not get renormalized. 

At low energies the sole contribution of physics between $\mu = \mw$ and 
$\mu = m_e$ therefore is to the running of the coupling $a_{klee}$ 
(and of the electric charge, $e$) between these scales, to all orders in
all other SM couplings. 

\vfill\eject
\subsection{Matching at $m_e$}
Next integrate out the electron itself to obtain the effective theory of photons and neutrinos only. At this stage we keep effective interactions having more than the minimal dimension, because these receive their largest coefficients when the
lightest possible particle -- the electron -- is integrated out. We obtain in this way all contributions to low-energy neutrino-photon physics which are
$O(1/\mw^2 m_e^p)$, for all $p$. 

To this order the new contributions to the effective neutrino-photon interaction 
lagrangian obtained by matching across the electron mass threshold is therefore given by
\eq
\label\nuphmaster
\Scl_{\rm elth}(\mu = m_e) =  {\GF \over \sqrt 2} \; \sum_{kl} 
\Bigl( i\nubar_k \gamma_\mu \Pl \nu_l \Bigr) \; \Bigl( v_{klee} 
\Avg{\ebar \, \gamma^\mu \, e} + a_{klee} \Avg{\ebar \, \gamma^\mu \gamma_5 e}
\Bigr), 
\eeq
where $\Avg{X^\mu}$ represents the expectation of the operator $X^\mu$, obtained by
integrating out the electrons, weighted by the QED lagrangian:\foot\reallymatch{A
notational aside is in order here, since eq.~\nuphmaster\ gives the impression that
$\ss \Avg{X^\mu}$ does not involve an integration over the electromagnetic field as well
as the electron field. In reality this expectation denotes the usual matching procedure:
the difference between the average calculated with electrons and photons in the theory
just above $\ss m_e$, and the average calculated with photons only in the effective
theory just below $\ss m_e$. This distinction plays no role in the present discussion.}
\eq
\label\eavg
\Avg{X^\mu} = \int \Scd e \Scd \ebar \; X^\mu(e,\ebar) \; \exp\left[ 
i \int d^4x \; \Bigl(\Scl_{\rm kin} -ie \, A^\mu \, \ebar \, \gamma^\mu \, e
\Bigr) \right] . 
\eeq

Eqs.~\nuphmaster\ and \eavg\ contain the nub of the main results, because it permits the following two conclusions:

\topic{Suppression of $2\to 2$ Processes}

As is easy to show, all operators involving only $\nubar \gamma^\mu\gamma_L \nu$ -- as 
opposed to $\nubar \gamma_\alpha \Pl \lrderiv{\beta} \nu$ -- and
two electromagnetic fields vanish on using the equations of motion for the neutrino
and photon fields, and so are redundant in the sense that they may be removed by
performing a field redefinition.  The only
possible lowest dimension operator for $2\to 2$ processes (dimension 6)
turns out to be of the form of eq.~\nunuphph.

It remains to show that operators of this form are always suppressed by at least
two powers of $\GF$. We have just argued that the right-hand-side of eq.~\nuphmaster\ is 
explicitly proportional to the neutrino current, $\nubar_k \gamma_\mu \Pl \nu_l$, and
so cannot contribute to an operator with a derivative embedded within the neutrino
bilinear. This is why integrating out the electron does not generate 
the operator, eq.~\nunuphph, with a coefficient proportional to $\GF/m_e^2$.
The same argument also precludes
generating such a term when the other charged leptons are integrated out. Charged-current 
interactions of neutrinos with quarks, on the other hand, can be linear in $\nu$,
and so need not be proportional to $\nubar_k \gamma_\mu \Pl \nu_l$. Nonetheless,
conservation of quark flavour only permits these interactions to contribute to
neutrino/photon scattering at second order in $\GF$.

\topic{Connection with Photon-Photon Scattering}

Since the electromagnetic interactions preserve parity ($\Scp$) and charge conjugation
($\Scc$), these symmetries may be used to further organize the contributions to
$\Scl_{\rm elth}$. In particular, these symmetries imply that any term in $\Scl_{\rm elth}$
involving an odd power of $F_{\mu\nu}$ receives contributions only from the vector
current, $\Avg{\ebar \gamma^\mu \, e}$, while those involving even powers of $F_{\mu\nu}$
arise purely from the axial current, $\Avg{\ebar \gamma^\mu \gamma_5 \, e}$. 

Furthermore, since the vector current, $\ebar \gamma^\mu \, e$, is also the electromagnetic
current for the electron effective theory, its expectation may be expressed in terms
of the Euler-Heisenberg effective lagrangian, $W_{\sss EH}[A]$, for photon-photon 
scattering below $m_e$ \Dicuslow, \Matiasone:
\eq
\label\EHconn
\eqalign{
\Avg{\ebar \gamma^\mu \, e} &= {1\over e} \; \left({\delta Z \over \delta A_\mu}\right), \cr
\hbox{where} \qquad
Z[A] &= e^{i W_{\sss EH}[A]} = \int \Scd e \Scd \ebar \; \exp\left[ 
i \int d^4x \; \Bigl(\Scl_{\rm kin} -ie \, A^\mu \, \ebar \, \gamma^\mu \, e
\Bigr) \right] . \cr}
\eeq

\ref\EH{H. Euler, {\it Ann.\ Phys. (Leipzig)} {\bf 26}, 398 (1936) \semi
W. Heisenberg and H. Euler, {\it Zeit.\ Phys.} {\bf 98}, 714 (1936).}

For instance, since the quartic contribution to the Euler-Heisenberg interaction
is given by: \EH
\eq
\label\EHfour
\Scl_{\sss EH}^{(4)} = {\alpha^2 \over 180 \, m_e^4}\;
\left[ 5 (F_{\mu \nu} F^{\mu \nu})^2 - 14 F_{\mu \nu}F^{\nu \lambda}
F_{\lambda \rho} F^{\rho \mu} \right] ,
\eeq
it follows that the dominant contribution to $\Scl_{\rm elth}$ involving two
neutrinos and three electromagnetic fields must be \Dicuslow, \Matiasone:
\eq
\label\effacttwo
\Scl^{(3)}_{\rm elth} = {e \, (\hf + 2\sw^2) \, \alpha \over 90 \pi \, m_e^4} 
\; \left( {\GF \over \sqrt2} \right) \Bigl[ 5 \, (N_{\mu\nu} \, F^{\mu\nu})
(F_{\lambda\rho} \, F^{\lambda\rho}) - 14 \, (N_{\mu\nu} \, F^{\nu\lambda}
\, F_{\lambda\rho} \, F^{\rho\mu} )\Bigr] ,
\eeq
with $N_{\alpha\beta} = \partial_\alpha \Bigl( \nubar \gamma_\beta \Pl \nu \Bigr)
- (\alpha \leftrightarrow \beta)$.

This method clearly works in general: to obtain any term
involving an odd power of $F_{\mu\nu}$ in $\Scl_{\rm elth}$,
replace one power of $e$ by 
$\GF \left(\hf+2 \sw^2\right)/\sqrt{2}$, and sum all possible ways of 
replacing one electromagnetic field strength by $N_{\mu\nu}$.

\subsection{The $2 \to 5$ Effective Lagrangian}

\def\outstate{\bra{\gamma_1\cdots\gamma_5}}

\ref\DicusEH{D.A. Dicus, C. Kao and W. Repko, \
\prd{57}{98}{2443--2447} ({\tt hep-ph/9709415}).}

The next simplest neutrino-photon interaction which is related in this way to
the Euler-Heisenberg action describes $2 \to 5$ processes, like $\nubar \nu
\to 5\gamma$. It is related to the sixth order term of the EH action, which
is given by \DicusEH:
\eq
\label\Leh
\Scl_{\ss EM}^{(6)} = {\pi \alpha^3 \over 315 \; m_e^8}
\left[9 (F_{\alpha \beta}F^{\alpha \beta})^3 -
26 F_{\mu \nu} F^{\nu \lambda} F_{\lambda \rho} F^{\rho \mu} 
\; (F_{\alpha \beta} F^{\alpha \beta}) \right] .
\eeq

Now, given our previous arguments 
we can read off the effective two-neutrino/five-photon operators directly.
We find:
\eq
\label\nugammafive
\eqalign{
\Scl^{\nu-5\gamma}_{\rm eff} &=  \; 
{\pi \over 315}\; {\alpha^{5 / 2} \over \sqrt{4 \pi}}\;
{\GF \over \sqrt{2}}\; {{\left({1 \over 2} + 2\sw^2\right)} \over m_e^8} 
\Bigl[ 6\cdot 9 (F_{\alpha \beta}F^{\alpha \beta})^2 F_{\mu \nu} N^{\mu \nu} \cr
& - 4\cdot 26 F_{\mu \nu} F^{\nu \lambda} F_{\lambda \rho} N^{\rho \mu} 
\; (F_{\alpha \beta} F^{\alpha \beta})  -
2\cdot 26 F_{\mu \nu} F^{\nu \lambda} F_{\lambda \rho} F^{\rho \mu} 
\; (F_{\alpha \beta} N^{\alpha \beta}) \Bigr],
}
\eeq

\ref\axialrefs{C. Schubert, {\tt hep-ph/0002276}; {\tt hep-ph/0001288}.} 

\ref\axiallag{H. Gies and R. Shaisultanov, {\tt hep-ph/0003144}.}

A similar method will also give the even powers of $F_{\mu\nu}$ in
$\Scl_{\rm elth}$ given the expression for the axial-vector/vector current
correlation in QED \foot\axialfoot{ After completing this paper it was
brought to our attention that, in ref.  \axiallag, the expression for
$\ss \avg{\ebar \gamma^\mu\gamma_5 e}$ has been worked out, up to fourth
order in the fields.  We thank H.~Gies for pointing this reference out to
us.}. 

\vfill\eject
\section{ Cross sections for ${\ol \nu}\; \nu \to 5\;\gamma$ and 
crossed processes}

\def\T#1#2{ T^{#1_1 #2_1,\cdots,#1_5 #2_5} }
\def\Ttwid#1#2{{\widetilde T}^{#1_1 #2_1, \cdots, #1_5 #2_5} }
\def\Fd#1#2{\partial_{#1} A_{#2}}
\def\Fds#1#2{\Fd{#1_1}{#2_1} \cdots \Fd{#1_5}{#2_5}}
\def\ke#1#2#3{{k_{(#1) \;#2} \epsilon_{#3}(k_{#1};\lambda_{#1}) \over 
\sqrt{(2 \pi)^3 k_{#1}^0}}}
\def\Ntwid{\widetilde N}

\ref\FORM{J.A.M. Vermaseren, KEK-Preprint-92-1; {\tt ftp://nikhefh.nikhef.nl}}

We next apply the lagrangian, eq.~\nugammafive, to compute the $2\to 5$ processes 
$\nubar \nu \to 5 \gamma$, $\nu \gamma \to \nu + 4\gamma$ and $\gamma \gamma
\to \nubar \nu + 3 \gamma$. This is a straightforward, if tedious, exercise within the
effective theory, requiring only the Born approximation using interaction
\nugammafive. (This should be contrasted with the difficulty of extracting the
low-energy limit of the scattering amplitude, computed directly from the
higher-loop graphs involving the weak interaction, eq.~\FFGMlagr, and QED.
As is usually the case with effective lagrangians, the payoff in simplicity
is much larger for nonleading contributions.)
In performing this calculation we employed the symbolic manipulation
program FORM \FORM, which reduced the squared amplitude
to its final form in under ten minutes on a desktop
PC.  We briefly sketch the method of computation below.

We require, then, the matrix elements of the effective interaction,
eq.~\nugammafive, which we write as follows:
\eqnn
\Scl^{\nu-5\gamma}_\eff = 
g N^{\mu \nu} \T{\alpha}{\beta}_{\mu \nu} \Fds{\alpha}{\beta}
\eeq
where $g$ is the factor premultiplying the square bracket in 
eq.~\nugammafive\ and $\T{\alpha}{\beta}_{\mu \nu}$ represents the
polynomial of momenta in the effective interaction. In terms of these
quantities the matrix element relevant to $\nubar\nu \to 5\gamma$, for
instance, becomes:
\eq
\outstate \Scl^{\nu-5\gamma}_\eff \ket{\nubar \nu} = 
g \Ntwid^{\mu \nu} 
\Ttwid{\alpha}{\beta}_{\mu \nu} 
 \prod_{i=1}^5\left(\ke{i}{\alpha_i}{\beta_i}\right), 
\eeq
where $\Ntwid^{\mu \nu} = \vacbra N^{\mu \nu} \ket{\nubar \nu}$
and
\eq
\Ttwid{\alpha}{\beta}_{\mu \nu} = \sum_{\pi \in S_5(1,\cdots,5)}
T_{\mu \nu}^{\alpha_{\pi_1} \beta_{\pi_1},
\cdots,
\alpha_{\pi_5} \beta_{\pi_5}} ,
\eeq
is the permutation-summed tensor contracting the fields together.
The $\epsilon^\mu$'s are, as usual, the photon polarisation vectors.

\def\dk#1{{\ddx{3}{k_#1} \over 2 k_#1^0}\;}

After squaring and doing the spin sums, the following phase-space
integral is required to obtain the total cross section:
\eq
\label\holygrail
\Sci_m^{\alpha_1 \ldots \alpha_m; \gamma_1 \cdots \gamma_m}(w) = 
\int\dk{1}\cdots \dk{m}k_1^{\alpha_1}k_1^{\gamma_1} \cdots 
k_m^{\alpha_m}k_m^{\gamma_m}\;\delta^4(\sum_{i=1}^m k_i - w).
\eeq
A general technique for evaluating integrals of this form is
given in the Appendix.

\topic{$\nubar\nu \to 5 \gamma$}
Using the integrals of the Appendix gives the final result for 
$\nubar\nu \to 5 \gamma$:
\eq
\eqalign{
\sigma(\nu {\ol \nu} \to 5 \gamma) & = 
{ 1487 \over (2 \pi)^6 2^4 3^9 5^4 7^4} \;
\alpha^5 m_e^2 \GF^2 \left({1\over 2}+2\sw^2 \right)^2 
\left({s \over  m_e^2}\right)^9 \cr
& \simeq 6.87 \pwr{-38} \left({s \over  m_e^2}\right)^9
\,{\rm barn},\cr
}
\eeq
where $s = - (p + \pbar)^2 = -2 p \cdot \pbar$ is the usual Mandelstam
variable, equal to $s = 4 E_\nu^2$ in the centre-of-mass frame. 

\topic{$\nu \gamma \to \nu + 4 \gamma$}
A similar exercise, after crossing the external lines, yields
\eq
\eqalign{
\sigma(\nu \gamma \to \nu + 4\gamma)
&= \sigma({\ol \nu} \gamma \to \nubar +  4\gamma) \cr
&= {13\cdot 163\cdot 2339 \over (2 \pi)^6\; 2^{8} 3^{10} 5^5 7^4 11 \;}
\alpha^5 m_e^2 \GF^2 \left({1 \over 2} +2 \sw^2 \right)^2 \; 
\left( {s\over  m_e^2 } \right)^9 \cr
&\simeq 8.68 \pwr{-38} \left( {s \over m_e^2 } \right)^9
\,{\rm barn}.
}
\eeq

\topic{$\gamma\gamma \to \nubar\nu + 3\gamma$}
Crossing the other neutrino yields,
\eq
\eqalign{
\sigma(\gamma\gamma \to {\ol \nu}\nu + 3\gamma) &= 
{797549 \over (2 \pi)^6\; 2^7 3^9 5^5 7^4 11} \;
\alpha^5 m_e^2 \GF^2 \left({1 \over 2} + 2\sw^2 \right)^2  
\left({ s \over m_e^2} \right)^9 \cr
&\simeq  8.38 \pwr{-38} \left({ s\over  m_e^2} \right)^9
\,{\rm barn}.
}
\eeq

\section{Conclusion}

We have constructed the effective interaction which governs the
interactions of five photons and two neutrinos using
the general connection between the
effective action for neutrino-photon interactions 
at lowest order in $\GF$ and the known
Euler-Heisenberg effective interaction for photon-photon scattering.  
As an application we have computed the two-body cross sections whose
low-energy limits are given in terms of this effective interaction.  While 
these cross sections are likely to be too small to
be of any astrophysical relevance, it is interesting that
the effective interaction for such a high-order process can
be obtained with such minimal effort.  It is also noteworthy that
a seventh-order process can compete with the two-body scattering
$\nubar \nu \to 2\gamma$.
Finally, our expressions were obtained by evaluating multi-body
phase space integrals, for which we have presented an efficient 
method of computation.

\section{Acknowledgements}

We would like to thank G. Mahlon and J. Matias for helpful 
conversations. This research was supported by funds from 
NSERC of Canada and FCAR of Qu\'ebec.

\appendix{I}{Phase Space Integrals}

\def\xtom#1{\left(x^2\right)^#1}
\def\eps{\epsilon}

Our goal in this appendix is to describe how to evaluate the integral, eq.~\holygrail,
which we reproduce once more for convenience:
\eqnn
\Sci_m^{\alpha_1 \ldots \alpha_m; \gamma_1 \cdots \gamma_m}(w) = 
\int\dk{1}\cdots \dk{m}k_1^{\alpha_1}k_1^{\gamma_1} \cdots 
k_m^{\alpha_m}k_m^{\gamma_m}\;\delta^4(\sum_{i=1}^m k_i - w).
\eeq
We will find, through the integral representation 
of the delta function, that we are able to reduce the problem to one of
taking derivatives of a suitable (simple) integral.
Since the general expressions are extremely lengthy, and 
particular results are not difficult to obtain with the help 
of a symbol manipulation program once the prescription is known, 
we will only provide a detailed recipe for evaluating these integrals.

We proceed by using the Fourier representation of the delta function to
factorise this integral into products of terms having the form
\eqnn
J^{\alpha \beta}(x) = \int {\ddx{3}{k} \over 2 k^0} 
k^\alpha k^\beta e^{ik\cdot x},
\eeq
so that
\eq
\label\Jint
\Sci_m^{\alpha_1 \ldots \alpha_m; \gamma_1 \cdots \gamma_m}(w) =  
\int\ddp{4}{x} J^{\alpha_1 \gamma_1}(x)\cdots J^{\alpha_m 
\gamma_m}(x) e^{-iw\cdot x}. 
\eeq

The integral defining $J^{\alpha\beta}$ is easily performed as follows:
\eq
\label\Jintdiff
J^{\alpha \beta}(x) = {1 \over (i)^2} {\partial \over \partial x_\alpha}
{\partial \over \partial x_\beta} \int {\ddx{3}{k} \over 2 k^0}\; e^{ik\cdot x} 
= {4 \pi \over x^6} (\eta^{\alpha \beta} x^2 - 4 x^\alpha x^\beta),
\eeq
where we use the integral
\eq
\int {\ddx{3}{k} \over 2 k^0} \; e^{ik\cdot x} = {2 \pi \over x^2} .
\eeq
We have ensured the convergence of this integral through the appropriate
$\epsilon$ prescription, taking
$x^2 = -(x^0)^2 + {\bf x}^2 + i\; sgn(x^0) \epsilon$.  The $\epsilon$ 
term in $x^2$ forces the incoming momentum to be future-pointing, 
and allows the $x$ integrals to be done unambiguously.  
Notice that our result for $J^{\alpha\beta}$ is traceless, 
as is required, since the $k$s are null.

With $J^{\alpha\beta}$ in hand, expand the integrand of~\Jint\ to obtain 
a sum of integrals of the form 
\eqnn
\int \ddx{4}{x} x^{\alpha_1} \cdots x^{\alpha_n} 
{e^{-i w\cdot x} \over (x^2)^m},
\eeq
where $w$ is a future-pointed, timelike four-vector.
These integrals can all be evaluated by differentiating
\eq
\label\easygrail
\Sci_m(w) \df \int \ddx{4}{x} {e^{-i w\cdot x} \over \xtom{m} },
\eeq
with respect to $w$, so that finding this integral reduces the problem
of calculating~\holygrail\ to one of expanding a polynomial and
taking derivatives.

\def\intR{\int_{-\infty}^{\infty}}
\def\intt{\intR dt\;}
\def\intr{\intR dr\; r^2}

Let us evaluate~\easygrail.  If we explicitly factor 
the denominator, go to the rest frame of $w$, and
consider the $t$ integral first, we need to consider
\eqnn
\label\Im
\eqalign{
\Sci_m(w) & = \int d^2\Omega \int_0^\infty dr r^2 \int_{-\infty}^{\infty} dt
{e^{+i \omega t} \over [\;-(t-i\eps)+r]^m[\;(t-i\eps)+r]^m} \cr
& = {1\over 2} \int d^2\Omega \intr \intt
{e^{+i \omega t} \over (-)^m[\;t-(r+i\eps)]^m[\;t+(r-i\eps)]^m} \cr
& = 2\pi (-)^m \intr \intt
{e^{+i \omega t} \over [\;t-(r+i\eps)]^m[\;t-(-r+i\eps)]^m}.\cr
}
\eeq
The $t$ integral in this expression 
is a contour integral, which is nonzero only for $\omega>0$, where
$\omega = w^0$.
For $\omega>0$ we close the contour upwards in a semicircle, break it into
two, one around each pole, and use the Cauchy integral formula to get
\eqnn
\eqalign{
& \Sci_m(w) = \intt {e^{+i \omega t} \over 
  [\;t-(r+i\eps)]^m[\;t-(-r+i\eps )]^m} \cr
& = {2\pi i \over (m-1)!}\left\{ 
{d^{m-1} \over dz^{m-1}}\left({e^{i z \omega} 
\over [z-(i\eps-r)]^m}\right)_{z=r+i\eps}
\!\!\!\! + 
{d^{m-1} \over dz^{m-1}}\left({e^{i z \omega} 
\over [z-(r+i\eps)]^m}\right)_{z=-r+i\eps}
\right\} \theta(\omega).
}
\eeq

Using the Leibniz rule, 
$d^m/dz^m(f(z)\;g(z)) = \sum_{s=0}^n 
 {m \choose s} f^{(n-s)}(z) g^{(s)}(z)$,
yields,
\eq
\label\tint
\eqalign{
\Sci_m(w) = & \;{2 \pi i \over [(m-1)!]^2 \; 2^m}
\sum_{s=0}^{m-1} {m-1 \choose s}\; (-)^s \;(m+s-1)! \cr 
& \times {(i \omega)^{m-1-s} \over 2^s}  \;
\left[ {e^{i r \omega} \over r^{m+s}} + 
(-)^{m+s}\; {e^{-i r \omega} \over r^{m+s}} \right] \theta(\omega).
}
\eeq
Now using this in eq.~\Im\ yields, after letting $r \leftrightarrow -r$ in the
second term of \tint,
\eq
\label\Imtwo
\eqalign{
\Sci_m(w) = & \; {8 \pi^2 i \over [(m-1)!]^2 \; 2^m}
\sum_{s=0}^{m-1} {m-1 \choose s}\; (-)^{m-s} \;(m+s-1)! \cr 
& \times \theta(\omega) {(i \omega)^{m-1-s}
\over 2^s} P \intR dr\; {e^{i r \omega} \over r^{m+s-2}}.
}
\eeq

Finally, since
\eqnn
P \intR dr \;{e^{i r \omega} \over r^n}
= {i \pi (i\omega)^{n-1} \over (n-1)!},
\eeq
substitution into \Imtwo\ yields
\eqnn
\eqalign{
\Sci_m(w) =&\; (-)^{m+1}\theta(\omega)\;{8 \pi^3  (i\omega)^{2m-4}\over [(m-1)!]^2 \; 2^m}
\sum_{s=0}^{m-1} {m-1 \choose s}\; {(-)^s\over2^s} \; 
(m+s-1)(m+s-2).
}
\eeq
The sum is easily recognised as the second derivative of
$2^{m-3}\,[x(1-x)]^{m-1}$, evaluated at $x=\hf$, and equal to ${-2(m-1) \over 2^{m-1}}$, so we obtain the general expression,
\eqnn
\Sci_m(w) =  \theta(\omega)\;{32 \pi^3 (m-1) \over [(m-1)!]^2 \; 4^m} (-w^2)^{m-2},\; 
{\rm for}\; m \geq 3.
\eeq

So, to summarise, by replacing the delta function by an integral, we are able to
do each of the $k$ integrals separately, obtaining~\Jint.  Collecting the
terms in the expansion and using~\easygrail\ and its derivatives to
proceed with the $x$ integrals yields the final answer to~\holygrail.

\vfill \eject

\listrefs

\bye

%% file: macros.tex

\font\titlefont = cmr10 scaled\magstep 4
 2
\font\sectionfont = cmr10
\font\littlefont = cmr5 
\font\eightrm = cmr8 

\def\ss{\scriptstyle} 
\def\sss{\scriptscriptstyle} 

\newcount\tcflag
\tcflag = 0  

\ifnum\tcflag = 0 \magnification = 1200 \fi  

\global\baselineskip = 1.2\baselineskip 
\global\parskip = 4pt plus 0.3pt 
\global\abovedisplayskip = 18pt plus3pt minus9pt
\global\belowdisplayskip = 18pt plus3pt minus9pt
\global\abovedisplayshortskip = 6pt plus3pt
\global\belowdisplayshortskip = 6pt plus3pt

\def\barsoff{\overfullrule=0pt}


\def\endignore{}
\def\ignore #1\endignore{} 

\newcount\dflag
\dflag = 0


\def\monthname{\ifcase\month 
\or January \or February \or March \or April \or May \or June%
\or July \or August \or September \or October \or November %
\or December 
\fi}

\newcount\dummy
\newcount\minute  
\newcount\hour
\newcount\localtime
\newcount\localday
\localtime = \time
\localday = \day

\def\advanceclock#1#2{ 
\dummy = #1
\multiply\dummy by 60
\advance\dummy by #2
\advance\localtime by \dummy
\ifnum\localtime > 1440 
\advance\localtime by -1440
\advance\localday by 1
\fi}

\def\settime{{\dummy = \localtime %
\divide\dummy by 60%
\hour = \dummy 
\minute = \localtime%
\multiply\dummy by 60%
\advance\minute by -\dummy 
\ifnum\minute < 10 
\xdef\spacer{0} 
\else \xdef\spacer{} 
\fi %
\ifnum\hour < 12 
\xdef\ampm{a.m.} 
\else 
\xdef\ampm{p.m.} 
\advance\hour by -12 %
\fi %
\ifnum\hour = 0 \hour = 12 \fi 
\xdef\timestring{\number\hour : \spacer \number\minute%
\thinspace \ampm}}}



\def\endtitle{}
\def\title#1\endtitle{\vskip.5in\titlefont
\global\baselineskip = 2\baselineskip 
#1\vskip.4in
\baselineskip = 0.5\baselineskip\rm}
 
\def\endauthors{}
\def\authors#1\endauthors{#1}

\def\endabstract{}
\def\abstract#1\endabstract{\vskip .3in%
\centerline{\sectionfont\bf Abstract}%
\vskip .1in
\noindent#1}

\def\nopageonenumber{\footline={\ifnum\pageno<2\hfil\else
\hss\tenrm\folio\hss\fi}}  

\newcount\nsection 
\newcount\nsubsection 

\def\section#1{\global\advance\nsection by 1
\nsubsection=0
\bigskip\noindent\centerline{\sectionfont \bf \number\nsection.\ #1}
\bigskip\rm\nobreak}

\def\subsection#1{\global\advance\nsubsection by 1
\bigskip\noindent\sectionfont \sl \number\nsection.\number\nsubsection)\
#1\bigskip\rm\nobreak}

\def\topic #1{{\medskip\noindent $\bullet$ \it #1:}}

\def\appendix#1#2{\bigskip\noindent%
\centerline{\sectionfont \bf Appendix #1.\ #2} 
\bigskip\rm\nobreak} 


\newcount\nref 
\global\nref = 1 

\def\therefs{} 

\def\semi{;\item{}} 

\def\ref#1#2{\xdef #1{[\number\nref]} 
\ifnum\nref = 1\global\xdef\therefs{\item{[\number\nref]} #2\ } 
\else
\global\xdef\oldrefs{\therefs}
\global\xdef\therefs{\oldrefs\vskip.1in\item{[\number\nref]} #2\ }%
\fi%
\global\advance\nref by 1
}

\def\listrefs{\vfill\eject\section{References}\therefs}


\newcount\nfoot 
\global\nfoot = 1 

\def\foot#1#2{\xdef #1{(\number\nfoot)} 
\hskip -0.2cm ${}^{\number\nfoot}$ 
\footnote{}{\vbox{\baselineskip=10pt
\eightrm \hskip -1cm ${}^{\number\nfoot}$ #2}}
\global\advance\nfoot by 1
}


\newcount\nfig 
\global\nfig = 1
\def\thefigs{} 

\def\figure#1#2{\xdef #1{(\number\nfig)}
\ifnum\nfig = 1\global\xdef\thefigs{\item{(\number\nfig)} #2\ }
\else
\global\xdef\oldfigs{\thefigs}
\global\xdef\thefigs{\oldfigs\vskip.1in\item{(\number\nfig)} #2\ }%
\fi%
\global\advance\nfig by 1 } 

\def\fig#1{\xdef #1{(\number\nfig)}
\global\advance\nfig by 1 } 


\newcount\ntab
\global\ntab = 1

\def\table#1{\xdef #1{\number\ntab}
\global\advance\ntab by 1 } 


\newcount\cflag
\newcount\nequation
\global\nequation = 1
\def\eqlabel{(1)}

\def\nexteqno{\ifnum\cflag = 0
\global\advance\nequation by 1
\fi
\global\cflag = 0
\xdef\eqlabel{(\number\nequation)}}

\def\lasteqno{\global\advance\nequation by -1
\xdef\eqlabel{(\number\nequation)}}

\def\label#1{\xdef #1{(\number\nequation)}
\ifnum\dflag = 1
{\escapechar = -1
\xdef\draftname{\littlefont\string#1}}
\fi}

\def\clabel#1#2{\xdef\eqlabel{(\number\nequation #2)}
\global\cflag = 1
\xdef #1{\eqlabel}
\ifnum\dflag = 1
{\escapechar = -1
\xdef\draftname{\string#1}}
\fi}

\def\cclabel#1#2{\xdef\eqlabel{#2)}
\global\cflag = 1
\xdef #1{\eqlabel}
\ifnum\dflag = 1
{\escapechar = -1
\xdef\draftname{\string#1}}
\fi}


\def\eeq{}

\def\eqnn #1\eeq{$$ #1 $$}

\def\eq #1\eeq{
\ifnum\dflag = 0
{\xdef\draftname{\ }}
\fi 
$$ #1
\eqno{\eqlabel \rlap{\ \draftname}} $$
\nexteqno}







\def\eqa #1\eeq{
\ifnum\dflag = 0
{\xdef\draftname{\ }}
\fi 
$$ \eqalignno{ #1 } $$
\global\cflag = 0}



\def\npb#1#2#3{{\it Nucl.\ Phys.} {\bf B#1} (19#2) #3}

\def\prd#1#2#3{{\it Phys.\ Rev.} {\bf D#1} (19#2) #3}
\def\pr#1#2#3{{\it Phys.\ Rev.} {\bf #1} (19#2) #3}

\def\prl#1#2#3{{\it Phys.\ Rev.\ Lett.} {\bf #1} (19#2) #3}


\global\nulldelimiterspace = 0pt


\def\df{\mathrel{:=}}


\def\frac#1#2{{{#1} \over {#2}}\,}  
\def\hf{{1\over 2}}



\def\Asl{\hbox{/\kern-.7500em\it A}} 
\def\Dsl{\hbox{/\kern-.6700em\it D}} 
\def\dsl{\hbox{/\kern-.5300em$\partial$}}
\def\pxpsl{\hbox{/\kern-.5600em$p$}}
\def\sslsh{\hbox{/\kern-.5300em$s$}}
\def\epssl{\hbox{/\kern-.5100em$\epsilon$}}
\def\delsl{\hbox{/\kern-.6300em$\nabla$}}
\def\lxpsl{\hbox{/\kern-.4300em$l$}}
\def\elxpsl{\hbox{/\kern-.4500em$\ell$}}
\def\kxpsl{\hbox{/\kern-.5100em$k$}}
\def\qxpsl{\hbox{/\kern-.5000em$q$}}
\def\sla#1{\raise.15ex\hbox{$/$}\kern-.57em #1}
\def\Pl{\gamma_{\sss L}}

\def\pwr#1{\cdot 10^{#1}}



\def\roughly#1{\mathrel{\raise.3ex\hbox{$#1$
\kern-.75em\lower1ex\hbox{$\sim$}}}}

\def\ol#1{\overline{#1}}





\def\Scc{{\cal C}}
\def\Scd{{\cal D}}

\def\Sci{{\cal I}}

\def\Scl{{\cal L}}

\def\Scp{{\cal P}}


\def\ssf{{\sss F}}

\def\ssl{{\sss L}}

\def\ssw{{\sss W}}

\def\ssz{{\sss Z}}


\def\pmb#1{\setbox0=\hbox{#1}%
\kern-.025em\copy0\kern-\wd0
\kern.05em\copy0\kern-\wd0
\kern-.025em\raise.0433em\box0}   


\font\jlgtenbrm=cmbx10
\font\jlgtenbit=cmmib10
\font\jlgtenbsy=cmbsy10
\font\jlgsevenbrm=cmbx10 at 7pt
\font\jlgsevenbsy=cmbsy10 at 7pt
\font\jlgsevenbit=cmmib10 at 7pt
\font\jlgfivebrm=cmbx10 at 5pt
\font\jlgfivebsy=cmbsy10 at 5pt
\font\jlgfivebit=cmmib10 at 5pt
\newfam\jlgbrm

\textfont\jlgbrm=\jlgtenbrm
\scriptfont\jlgbrm=\jlgsevenbrm
\scriptscriptfont\jlgbrm=\jlgfivebrm
\newfam\jlgbit

\textfont\jlgbit=\jlgtenbit
\scriptfont\jlgbit=\jlgsevenbit
\scriptscriptfont\jlgbit=\jlgfivebit
\newfam\jlgbsy

\textfont\jlgbsy=\jlgtenbsy
\scriptfont\jlgbsy=\jlgsevenbsy
\scriptscriptfont\jlgbsy=\jlgfivebsy
\newcount\jlgcode
\newcount\jlgfam
\newcount\jlgchar
\newcount\jlgtmp
\def\bolded#1{
        \jlgcode\the#1 \divide\jlgcode by 4096
        \jlgtmp\the\jlgcode \multiply\jlgtmp by 4096
        \jlgfam\the#1 \advance\jlgfam by -\the\jlgtmp
        \divide\jlgfam by 256
        \jlgtmp\the\jlgcode \multiply\jlgtmp by 16
	\advance\jlgtmp by \the\jlgfam
	\multiply\jlgtmp by 256
        \jlgchar\the#1 \advance\jlgchar by -\the\jlgtmp
        \advance\jlgfam by \the\jlgbrm
        \jlgtmp\the\jlgcode
        \multiply\jlgtmp by 16
        \advance\jlgtmp by \the\jlgfam
        \multiply\jlgtmp by 256
        \advance\jlgtmp by \the\jlgchar
        \mathchar\the\jlgtmp
}


\def\Im{{\rm Im\;}}


\def\bra#1{\langle #1 |}
\def\ket#1{| #1 \rangle}

\def\avg#1{\langle #1 \rangle}

\def\Avg#1{\left\langle #1 \right\rangle}

\def\ddx#1#2{d^{#1}#2\,}
\def\ddp#1#2{\frac{d^{#1}#2}{(2\pi)^{#1}}\,}


\def\vacbra{{\bra 0}}
